\documentclass[aps,prl,reprint,citeautoscript, floatfix]{revtex4-1}

\usepackage{graphicx}
\usepackage{dcolumn}
\usepackage{bm}
\usepackage{braket}   
\usepackage[colorlinks, linkcolor=blue, citecolor=blue, urlcolor=blue]{hyperref}
\usepackage{url}

\usepackage{natbib}
\usepackage{datatool}
\usepackage{etoolbox}
\usepackage{xcolor}
\usepackage[normalem]{ulem}

\usepackage{float}

\newcommand{\miold}[1]{\iffalse{#1}\fi}

\DTLnewdb{bibnotes}

\begin{document}

\preprint{APS/123-QED}

\title{Characterization of Exciton-exciton entanglement and correlations}


\author{Fangzhou Zhao$^1$}
\affiliation{$^1$ Theory Department, Max Planck Institute for the Structure and Dynamics of Matter and Center for Free-Electron Laser Science, Hamburg 22761, Germany}
\author{Carlos Mejuto-Zaera$^2$}
\affiliation{$^2$ Univ Toulouse, CNRS, Laboratoire de Physique Th\'eorique, Toulouse, France.}
\author{Angel Rubio$^{1,3}$}
\affiliation{$^1$ Theory Department, Max Planck Institute for the Structure and Dynamics of Matter and Center for Free-Electron Laser Science, Hamburg 22761, Germany}
\affiliation{$^4$ Center for Computational Quantum Physics (CCQ), The Flatiron Institute, New York, New York 10010, USA}
\author{Vojt\v{e}ch Vl\v{c}ek $^{4,5}$}
\affiliation{$^4$ Department of Chemistry and Biochemistry, University of California, Santa Barbara, Santa Barbara, California 93117, USA}
\affiliation{$^5$Materials Department, University of California, Santa Barbara, Santa Barbara, California 93117, USA}

\date{\today}

\begin{abstract}
Excitons in the weakly interacting regime can be well-described by many-body perturbation theories such as the Bethe-Salpeter equation formalism. However, for materials such as transition metal dichalcogenides moiré heterostructures under strong illumination, with the emergence of dense excitonic states, the strong correlation and entanglement between electrons and holes may cause the many-body perturbation method to fail, and excitons may not be treated in the bosonic picture, but exhibit fermionic behaviors. 
In our work, we investigate the phase space where excitons, and the electrons and holes which constitute them, are weakly or strongly entangled, as well as their binding for different interaction profiles and the degree of localization of the electrons and holes.  We corroborate the validity of using many-body perturbation theory in the exciton with interactions. 
Our work provides a general way to analyze the correlation and entanglement of multi-particle excitations in many-body systems, and gives a more comprehensive understanding of different phases for exciton entanglement and interactions in 1D systems.  
\end{abstract}

\maketitle

\section{Introduction}

Understanding the interaction between excitons has proven instrumental for the description of many physical phenomena in condensed matter physics such as Bose-Einstein condensation of excitons \cite{blatt1962bose, eisenstein2004bose}, excitonic insulators \cite{jerome1967excitonic, cercellier2007evidence, lu2017zero, jia2022evidence}, topological excitonic insulators \cite{hossain2025topological}, Mott-superfluid transitions \cite{fisher1989boson, greiner2002quantum}, bosonic Mott insulators \cite{endres2011observation}, and cavity quantum electrodynamics \cite{noh2016quantum, carusotto2013quantum, ballarini2019polaritonics}. 
Recent experiments demonstrated strong interaction and correlations of excitons in moiré superlattices \cite{xiong2023correlated}. 
In these cases, it is common to employ, e.g., a Bose-Hubbard model~\cite{fisher1989boson, song2025solvable, takasu2020energy, yang2020observation} to describe the exciton-exciton, as well as the electron-exciton repulsion.
Models based on the Bose-Hubbard Hamiltonian using Lindblad equations have also been proposed to analyze the 
properties of exciton-exciton interaction phase transitions \cite{song2025solvable}.
However, comprised of interacting electron and hole quasiparticles, excitons do not form ideal bosonic particles \cite{laikhtman2007excitons, hanamura1970theory}. Further, they can, in principle, have
strong interactions with the electron and hole in another exciton, so a coarse-grained bosonic description may not capture their physics accurately. 
Consequently, microscopic models of exciton-exciton interactions have come forward, including explicit fermionic and dipole-dipole terms at the level of the Bethe-Salpeter equation (BSE) \cite{steinhoff2024exciton, cudazzo2020correlation}. 
Nevertheless, these typically start from a dilute exciton limit, in which strong correlations between the excitons themselves are neglected.

Understanding the exciton-exciton interactions and entanglement from first principles is more challenging since it formally involves solving 8-point Green's functions to capture the four-particle excitations.   
While the BSE formalism is arguably the most widely applied method to this end, doing so \emph{ab initio} requires assuming weak excitonic correlation and approximating the equations~ \cite{strinati1988application, rohlfing2000electron, rohlfing1998electron, qiu2013optical, naik2022intralayer, zhao2025excitonic}. 
Given the computational complexity of BSE calculations, evaluating exciton-exciton interaction by solving multi-exciton properties from first principles based on Green function theory is extremely difficult.

Based on models such as the Bose-Hubbard model, exciton-exciton interaction in dense exciton gases \cite{may1985many, boldt1985many}, and rich phases of correlated bosons have been studied using many-body perturbation theories (MBPT) \cite{kagan2002two, lo2016self, alhyder2025lattice}. 
However, more fundamentally, the applicability of MBPT to describe excitonic properties relies largely on the strength of exciton-exciton interactions and correlations, as well as the electron–hole correlation within a single exciton.
When the electron-hole interaction is weak, the excitonic wavefunction can be approximated as a direct product of quasi-electron and quasi-hole wavefunctions, and perturbative linearized equations are good approximations to describe excitonic properties. 
However, when the excitons are strongly correlated, or the electron and hole forming an exciton cannot be treated independently, their wavefunctions need to be described by many-body rather than quasiparticle wavefunctions, and excitons exhibit fermionic behavior.

In this work, to better understand the correlated exciton phases such as the excitonic Mott-insulator \cite{lian2024valley, gao2024excitonic} and exciton superfluid \cite{liu2022crossover, li2017excitonic}, as well as to explore in which regime the MBPT can be used to treat excitons. 
We investigate exciton-exciton interactions and characterize the strength of exciton-exciton correlation and entanglement into different phases, when the repulsion interaction between the electrons (holes) and electrons (holes), and the pairing field between electrons and holes, present different strengths and distance dependencies.

Four distinct scenarios can be identified, corresponding to when the pairing field, i.e., the electron-hole (e-h) attraction, dominates over the repulsive interactions at both short and long ranges (purely attractive ``PA"), only at short range (short-range attractive ``SA"), only at long range (long-range attractive ``LA"), or being weaker than the repulsive interactions at both ranges (purely repulsive ``PR").
We would expect that the excitons will be more correlated when the pairing field dominates at long range, and the exciton will be locally bound or confined when the pairing field dominates at short range.

We design a minimal model hosting these different regimes to analyze their degree of exciton-exciton correlation.
In order to apply exact approaches and to avoid artifacts due to unscreened interactions, we solve non-periodic finite-size instances of our model.
We extrapolate our results to the thermodynamic limit and 
analyze if the correlated nature of the many-body wavefunction can be downfolded to an effective few-body wavefunction.
The excitonic phase diagram will be plotted for different interaction strengths and distance dependencies.

\begin{figure}[b]
\includegraphics[width=86mm]{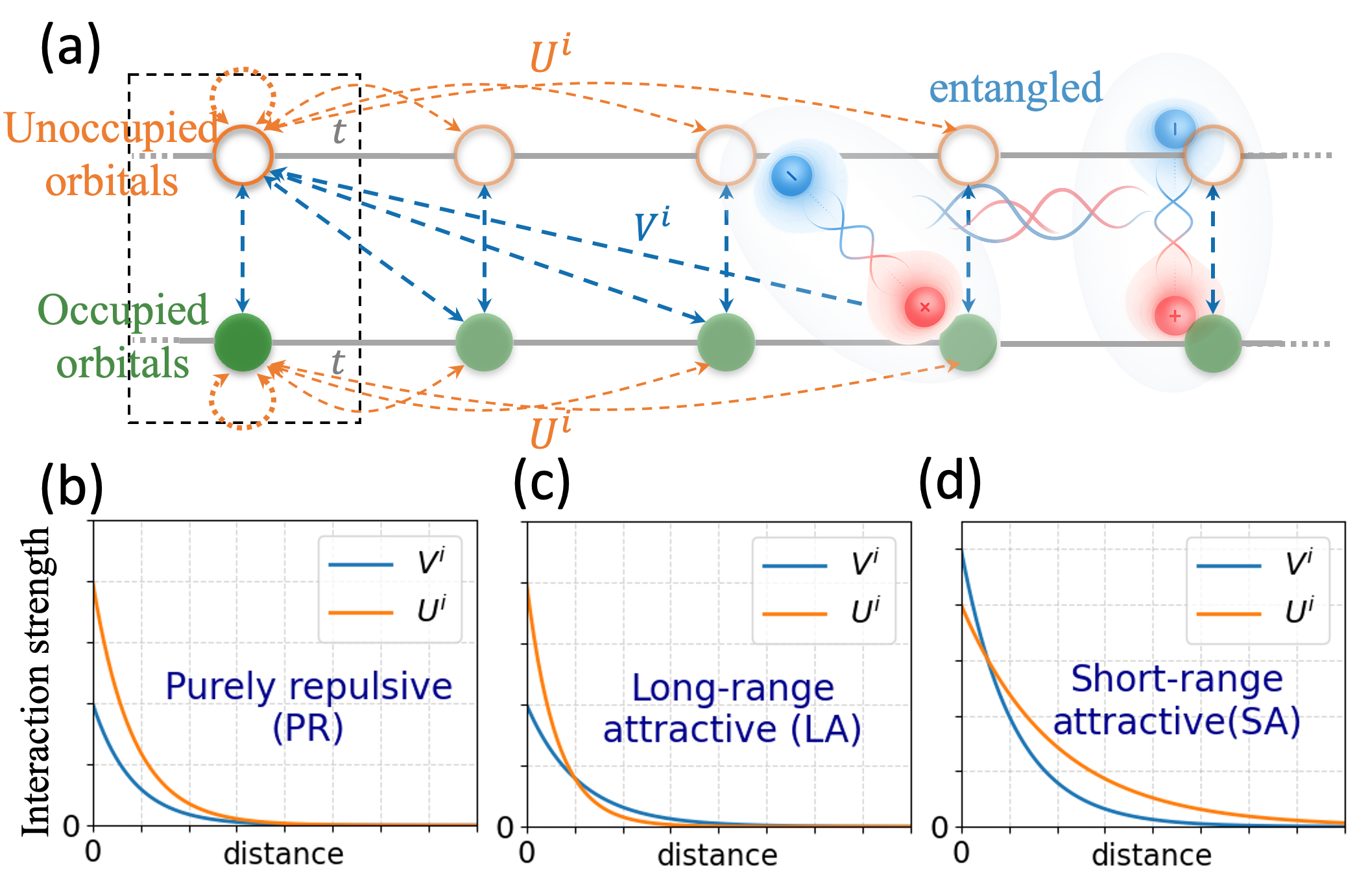}
\caption{\label{fig:1} (a) Our 1D model to investigate exciton-exciton interactions. (b-d) Exponential decay of the attractive and repulsive interactions.
(b) A case where the repulsive interaction dominates at all distances. (c) A case where the repulsive interaction dominates at shorter distances, but the attractive interaction dominates at longer distances.  (d) A case where the attractive interaction dominates at shorter distances, but the repulsive interaction dominates at longer distances. The quantum confinement of multiexcitations occurs in this case. 
}
\end{figure}

Our two-band model can be is inspired by systems hosting interlayer excitons, e.g., moiré systems \cite{xiong2023correlated, jin2019identification, naik2022intralayer, jin2019observation}. For computational simplicity, we focus here on a 1D realization of our model, which can be immediately relevant for quasi-1D materials with tunable electronic structure by doping or external field, such as single-wall carbon nanotubes with different chirality numbers  \cite{dresselhaus1998carbon, barros2006review}, moiré superlattice formed in double-wall carbon nanotubes \cite{zhao2022interlayer}, as well as doping, width, and electric field in graphene nanoribbons \cite{son2006energy, zhao2021topological}. 
We defer the study of 2D realizations to future work.
Our minimal two-band one-dimensional model contains long-range repulsive ($U$) and pairing ($V$) density-density interactions, illustrated in Fig.\ref{fig:1}(a). The pairing $V^i$ is restricted to be between electrons in conduction states and holes in the valence states, paralleled with interlayer excitons.
The hopping scaled by the onsite attraction  {\color{blue} ($t/V^0$)}is inversely related to the localization of the electrons and holes.
The magnitude and decay rate of $V^i$ and $U^i$  can also be tuned by different screening strengths in the material by different substrates. 
The Hamiltonian of our model reads: 
\begin{align}
H =\; & - \sum_{i=j , \langle i,j \rangle} t_{i, j}^c c_i^{c\dagger} c^c_j - 
     \sum_{i=j , \langle i,j \rangle} t_{i, j}^v d_i^{v\dagger} d^v_j + \text{c.c.} \\
& + \sum_{i, j} U_{ij} ( n_i n_{j} + p_i p_{j} )\notag  - \sum_{i, j} V_{ij} (p_i n_{j} + n_i p_{j})   ,
\end{align}
where  $c_i^{c\dagger}$ and $d_i^{v\dagger}$ are the creation operators for electron and hole on the unoccupied and occupied orbital on the $i$ th site, respectively. 
The occupation of electrons on the unoccupied orbital $n_i = c_i^{c\dagger} c^c_i$ and the occupation of holes on the occupied orbitals $p_i = d_i^{v\dagger} d^v_i$. 
When $i=j$, $t_{i,i}^c = \Delta$ and $t_{i, i}^v = -\Delta$ 
are the onsite energies, $t_{\langle i,j \rangle} = t$ is the hopping, where $\langle i,j \rangle$ denotes the nearest neighbors.
{ 
To explicitly study exciton-exciton interactions, we investigate doubly excited states of our model on top of a half-filled configuration, where the lower band is fully occupied, and the upper band is fully empty.
The states correspond to configurations containing two excitons and encompassing different types of electron–hole excitations, ranging from free particles to biexcitons, among others. 
{We solve the many-body wavefunctions of our model using exact diagonalization (ED) in the singly and doubly excited subspaces (see SM).}
{We mainly investigate the degree of correlation by changing $t/V_0$, i.e., the degree of localization, while keeping the onsite attraction $V^0$ as a constant with respect to the gap size, $V^0/ \Delta = 1/8 $, corresponding to a value estimated in the poly(\textit{p}-phenylene vinylene) polymer \cite{zhao2025excitonic} SM, S3).}

Although screening is reduced in low-dimensional materials, it still makes the decay of interactions exponential, so we assume geometrical decay of $V^i$ and $U^i$ with distance (rescaling 
rate defined as $V^{i+1} = \gamma_V V^{i}$, and $U^{i+1} = \gamma_U U^{i}$).
{With $\frac{U^0}{V^0} > 1$ or $< 1$, and $\frac{\gamma_U}{\gamma_V} > 1$ or $< 1$, the model falls in 4 different cases: PR [Fig. \ref{fig:1} (b)], LA [Fig. \ref{fig:1} (c)], SA [Fig. \ref{fig:1} (d)], and PA. 
However, we do not consider the PA case since the model does not exhibit e-h dissociation in this case because exciton formation will be energetically stabilizing already in the ground state (without excitation).}
{We expect that the LA regime will harbor highly correlated phases due to the long-range attraction, and the SA regime may give strong binding of excitons in smaller sizes. We will use the numerical result to verify our expectations. }

We will show that our model yields multiple phases characterized by doubly excited states. 
To distinguish between them, we define different levels of exciton binding using energetics compared to the fundamental gap $E_g$ (See SM, S5) and the optical gap $E_g^{opt}$.
$E_g^{opt}$ is defined as the lowest excitation energy $E^{n=1}_1$.
When there exists at least one double excitation state below double the optical gap $2E_g^{opt}$, there exists binding between excitons, so we define it as the strongly-bound (SB) phase [Fig. \ref{fig:2}(a)], which can be bi-exciton formation or formation of a trion with a charge-compensating electron or hole. When all the double excitation state energies are above $2E_g$, we define the case as the dissolved e-h or e-h plasma phase {[Fig. \ref{fig:2} (a)]. }
The boundary for the system to enter the SB phase and the dissolved phase are shown by the blue and pink square dots in Fig. \ref{fig:2}(b, c) for $2 \le N_{site} \le 9$.

To infer the phase boundaries at the thermodynamic limit $N_{site} = \infty$, we extrapolate 
the calculated 
phase boundaries in the $\frac{|t|}{|V^0|}$ vs $1/N_{site}$ plot. 
Since there are no known laws to describe this extrapolation, we used inverse power-law functions of different orders and used a gradient color to represent the uncertainty in the phase boundaries. 
In fact, we find that the geometrical decay of $V^i$ and $U^i$, which properly incorporates the screening, will lead to finite values in the thermodynamic limit, indicating that the model remains well defined. 

{In the PR and LA regime, the e-h dissociation occurs at larger hopping in larger system size [Fig. \ref{fig:2} (b)]. 
Interestingly, in the SA regime, a "quantum confinement effect" emerges
in regions of parameter space where the dominance of short-range attraction is strong and the system size is commensurate with the extent of the attractive region. 
As an example shown in Fig. \ref{fig:2} (c), where the system is in SA case, $t_{dis}$ shows non-monotonic behavior dependence on $N_{site}$. The e-h dissociation requires even higher hopping when $N_{sites}$ is small, 
with a minimum dissociation kinetic energy at $N_{sites} = 6$. 
The e-h are more tightly bound at a smaller size; in other words, it exhibits quantum confinement effects for e-h pairs, confirming our expectation that the SA regime would result in stronger exciton binding at smaller system sizes.}
It is physical since only when the excitons are localized in a small area, the short-range interactions shown in Fig. \ref{fig:1} (d) dominate the interaction between the excitons, which binds the excitons together.  
We further note that $t_{dis}$ shows perfect parabolic dependence on $1/N_{site}$.

We have also calculated the hopping threshold to dissolve all the single excitations from bound e-h pairs except the first excited state, and so far have not found any quantum confinement effect for single excitations (see SM), even when the near-field attraction is much stronger than the repulsions. 
This counterintuitive phenomenon suggests that interactions mediated through multi-exciton states are essential to confine the e-h pairs in small-size systems, which suggests that the quantum confinement for exciton states we discover is a many-body (multi-exciton) effect.

We turn to the main many-body effect for the present study, the degree of exciton correlation, in the thermodynamic limit.
We define it as follows:
We name the strongly correlated exciton phase  the exciton fluid, within which two sub-regimes can be identified. 
In the strongly correlated exciton fluid (SEF), both inter-exciton correlations and electron–hole correlations within each exciton are strong, such that the many-body wavefunction cannot be factorized into a direct product of excitons. 
In the weakly correlated exciton fluid (WEF), only the intra-exciton electron–hole correlations are strong, allowing the many-body wavefunction to be written as a direct product of excitons, but not of independent electrons and holes. 
For weakly correlated cases where the wavefunctions can be factored as a direct product of electrons and holes, we name the phase as exciton gas (EG). 
The description of the correlated phases is formalized using Green's functions in Appendix B.
In Fig.~\ref{fig:2} (b) and Fig.~\ref{fig:4}, we characterize the degree of correlation in terms of a many-body wavefunction projection method described in Appendix C.
In the Supplementary Material (SM), we present an equivalent characterization in terms of two-orbital entanglement.
We investigate the existence of exciton fluid (EF) phases (a combination of SEF and WEF phases) in the regime where the e-h pairs are moderately bound, 
i.e., the regimes outside the SB and e–h plasma phases, since in those phases, excitons may not be well-defined elementary excited quasiparticles. 
While the phase boundaries quantitatively depend on arbitrarily chosen threshold values for the wave function projection, we find the EF phases are mostly in the range $\sim 0.5 < \frac{t}{V^0} < 2$ for all meaningful thresholds.

\begin{figure*}[!t]
\includegraphics[width=180mm]{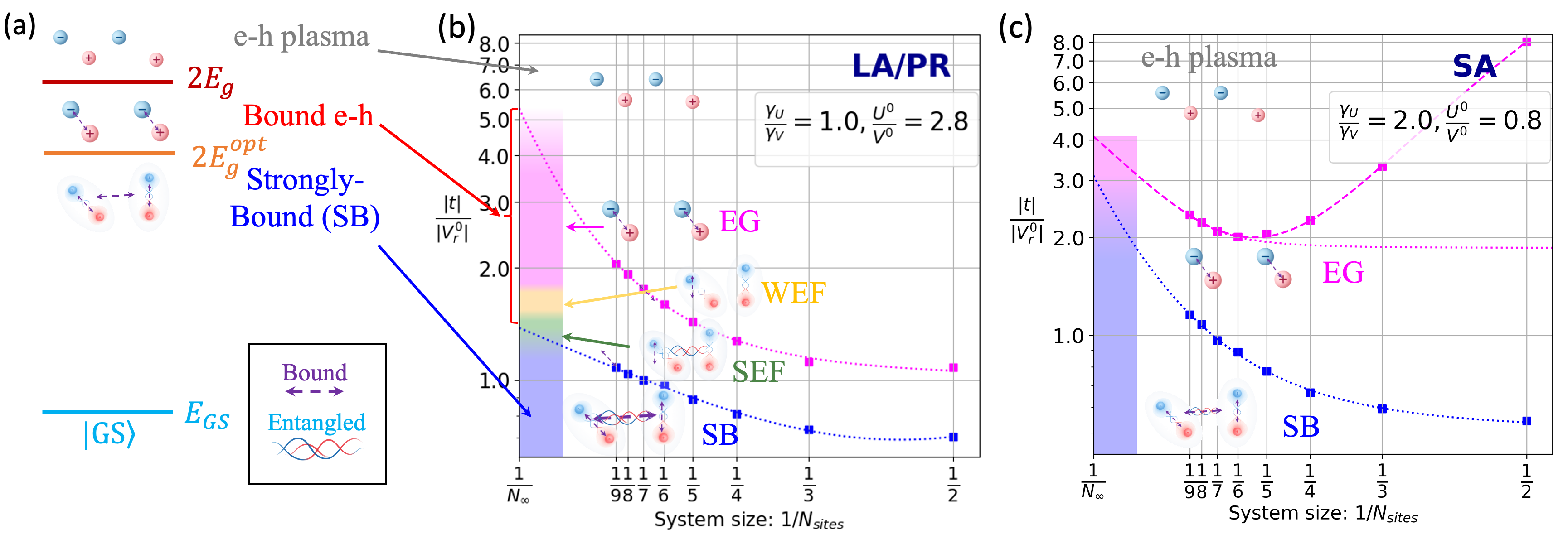}
\caption{\label{fig:SI_extra} 
Calculated phase boundaries and extrapolation of the thermodynamic limit of phase transitions. (a) Schematic diagram for double excitation states when they are characterized in e-h plasma, bound e-h, or strongly bound phase. The purple dashed arrows denote the binding between particles, and the blue and red wavy lines denote the entanglement. 
(b) When $\frac{\gamma_U}{\gamma_V} = 1.0, \frac{U^0}{V^0} = 2.8 $, it is in the boundary of an LA model and a PR model, corresponding to the purple linecut in Fig. \ref{fig:4}(b). In this case, both 
the boundary to enter SB and EG
are monotonously increasing as $N_{site}$ goes larger, so there is no quantum confinement effect. 
(c)When $ \frac{\gamma_U}{\gamma_V} =2.0 $, $ \frac{U^0}{V^0} = 0.8 $, it is an SA model, corresponding to the purple linecut in Fig. \ref{fig:4}(a). It shows the quantum confinement effect for multi-excitations. 
 \label{fig:2} 
 }
\end{figure*}

\begin{figure}[!h]
\includegraphics[width=85mm]{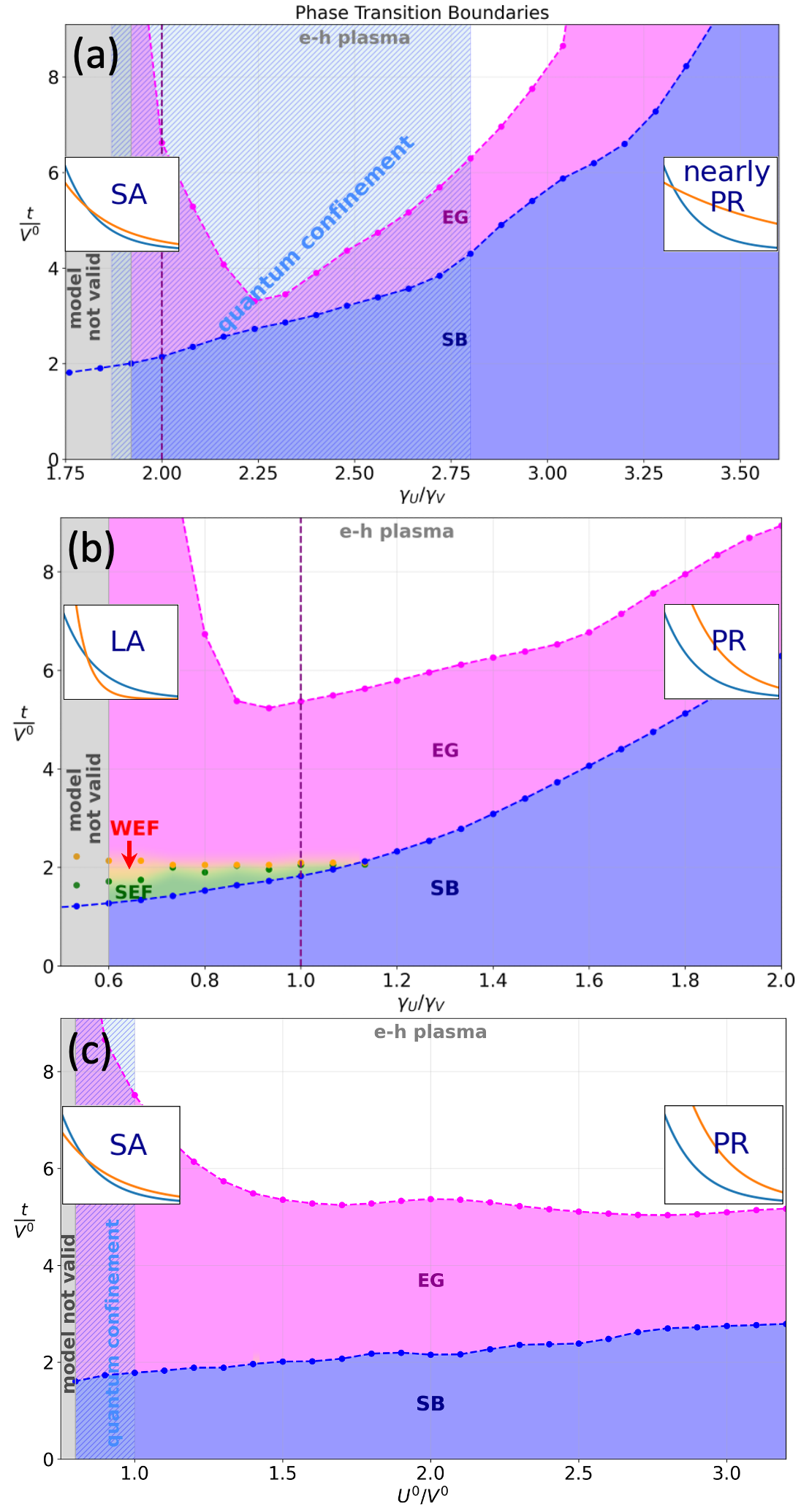}
\caption{\label{fig:4} 
Phase diagrams in the thermodynamic limit representing transitions between different cases. (a) Phase diagram for fixed $ U^0/V^0 = 0.8 $ and changing $ \gamma_U/\gamma_V $, the system goes from SA models to nearly PR models. The quantum confinement happens when $ \sim 1.9 < \gamma_U/\gamma_V < \sim 2.8$, where the model is SA. At large $ \gamma_U/\gamma_V $, both the phase boundary between SB and EG and the phase boundary between EG and e-h plasma exhibit quadratic behavior. The purple line corresponds to the case shown in Fig.\ref{fig:2}(c).
(b) Phase diagram for fixed $ U^0/V^0 = 2.8$, and changing $ \gamma_U/\gamma_V $, the system goes from LA models to PR models. The system has strongly correlated exciton phases when $ \gamma_U/\gamma_V < \sim 1.0$ where the model is LA. The purple line corresponds to the case shown in Fig.\ref{fig:2}(b).
(c) Phase diagram for fixed $ \gamma_U/\gamma_V = 2.0$, and changing $ U^0/V^0 $, the system goes from SA models to PR models. The system has quantum confinements when $ U^0/V^0 \sim 1.0 $ where the model is SA. At large $ U^0/V^0$, both the phase boundary between SB and EG and the phase boundary between EG and e-h are nearly constant. 
}
\end{figure}

{We investigate how the phase boundary evolves among different cases (SA, LA, and PR) of our model, by changing the ratio of the decay rate 
of the repulsive and attractive interactions $\frac{\gamma_U}{\gamma_V}$, and the ratio between the  onsite repulsive and attractive interactions $\frac{U^0}{V^0}$. }
We show three characteristic scenarios (Fig. \ref{fig:4}). 
In Fig. \ref{fig:4}(a), while fixing $U_0/V_0 = 0.8$, the model is formally in the SA regime when $\gamma_U/\gamma_V > 1$, where the exciton radius is commensurate with the pairing region, yet the exciton can still dissociate. 
But as $\gamma_U/\gamma_V$ increases to larger values, the attractive interaction only dominates over a very small distance, 
so we denote this case as "nearly PR". 
When $\sim 1.85 < \gamma_U/\gamma_V < \sim 2.8$, 
the pairing interaction dominates over a short range such that excitons are more stable when the system size is commensurate with such range, leading to quantum confinement. 
Across both the SA and PR regimes, we show the absence of a strongly correlated phase, and the SB regime directly transits to the EG regime.
When $ \gamma_U/\gamma_V < \sim 1.85 $, the thermodynamic limit of e-h dissociation energy diverges. 
This arises because the model is unphysical due to its proximity to the purely attractive regime.

In Fig.  \ref{fig:4}(b), while fixing $U_0/V_0 = 2.8$, $\gamma_U/\gamma_V < 1$ and $> 1$ are the LA and PR case respectively. 
In the PR regime, as the hopping increases, the system only goes from the SB phase to the EG phase, and then to the e-h plasma.  
But in the LA case, there exist the WEF and SEF strongly correlated excitonic phases in the regime where e-h pairs are moderately bound. 
The EF phase only exists when there is long-range attraction. 
This matches our expectation that long-range pairing field causes the strong correlation of excitons. 
And our extrapolation to the thermodynamic limit has successfully captured the correlations at the thermodynamic limit due to the long-range attractions. 

Both Fig. \ref{fig:4} (a) and (b) show that the dissociation of e-h occurs at lower hopping when $\gamma_U/\gamma_V$ increases, and increases to larger hoppings as $\gamma_U/\gamma_V$ further increases. At small $\gamma_U/\gamma_V$, increasing $\gamma_U/\gamma_V$ causes more repulsive interactions, so the e-h became easier to dissociate. The increase of the e-h dissociation boundary at larger $\gamma_U/\gamma_V$ seems counterintuitive. This arises from the fact that the large repulsive interactions in this range caused the wavefunction of the excitonic state to be more evenly distributed along the model chain, so the effective attractions between the e-h pairs effectively increase since the onsite e-h attraction exists on each site (see SM).
We also found that the wavefunction for the double excited state is more localized at larger  $\gamma_U/\gamma_V$. Larger localization causes stronger binding between the excitons, so the phase boundary between SB and EG increases monotonously in Fig. \ref{fig:4} (a) and (b).

Figure \ref{fig:4} (c) displays a complementary cut of the phase diagram, obtained by fixing $\gamma_U/\gamma_V = 2.0$ and varying $U_0/V_0$.
When $U_0/V_0 < 1$, the model is in an SA regime, and we observe that quantum confinement happens in a small region around $U_0/V_0 \sim 1$. The model becomes invalid when $U_0/V_0 < \sim 0.7$ since the strong short-range binding precludes e-h dissociation.  This further supports our conclusion that quantum confinement usually occurs in the SA model. 
In this case, the phase boundary between EG and e-h plasma remains nearly constant at larger $U_0/V_0$. The effect of a more evenly distributed wavefunction resulting in more onsite e-h attraction, and the effect of larger repulsion cancel each other in this case. In all parameter range, the strongly correlated exciton phases are absent. In our investigations, the EF phases only exist in the LA models.

In conclusion, we have used a minimal model to explore the possible types of exciton correlation in solids.
Our results have uncovered the potential of a quantum confinement effect for multi-exciton states.
The e-h pairs in multi-excitations can be stabilized in a smaller system size, 
which occurs when the attractive interaction dominates in the near-field range.
The confinement effect is a multi-exciton many-body effect. 
The quantum confinement for multi-excitation e-h pairs is inspiring since such confinement enhances the electron–hole wavefunction overlap, resulting in higher radiative recombination rates and higher binding energy, which provides guidance for designing quantum-dot-based light-emitting devices with high emission efficiency or prolonged exciton lifetimes.

We have discovered that in the parameter space where usually there is no attractive {net} interaction in long range, the system goes from the strongly bounded phase to the exciton gas phase, and then e-h pairs dissociate, as the electrons and holes go to more itinerant (large $t/V^0$) from localized (small $t/V^0$). 
But when the pairing dominates at long range, there are strongly correlated exciton fluid phases, where excitons exhibit fermionic behavior, between the strongly bound phase and the dissociation of e-h pairs. 
Over the most range of parameter space for the interactions, multi-exciton states lie in a weakly correlated or weakly entangled exciton-gas phase, if there are no extra bindings between excitons (bi-exciton or trions). 
Thus, many-body perturbation theories such as Bethe-Salpeter equations, which can be applied to weakly correlated systems, can be used to treat exciton states in the majority of the cases.

The phase boundaries we predicted between exciton fluid and the exciton gas are usually at $\sim 1 < t/V^0 < \sim 2$, which is almost independent of the relative strength of the repulsive and attractive interactions and their profiles.  
Thus, whether the electrons and holes are itinerant or localized determines the degree of correlation and entanglement of the excitons.

\begin{acknowledgments}

We acknowledge fruitful discussions with C. Fan. 
F. Z. acknowledges the support of the Alexander von Humboldt-Stiftung for the financial support from the Humboldt Research Fellowship.
This work was supported by the European Research Council (ERC-2024-SyG-101167294-UnMySt), the Cluster of Excellence ``CUI: Advanced Imaging of Matter"-EXC 2056-project ID 390715994.
We acknowledge support from the Max Planck-New York City Center for Non-Equilibrium Quantum Phenomena. The Flatiron Institute is a division of the Simons Foundation.
Computational resources were provided by the National Energy Research Scientific Computing Center, a DOE Office of Science User Facility supported by the Office of Science of the DOE under Contract No. DE-AC02-05CH11231. 
The collaborative work has been supported equally by the UC Santa Barbara NSF Quantum Foundry funded via the Q-AMASE-i program under award DMR-1906325 and Eddleman Quantum Institute at UC Santa Barbara.

\end{acknowledgments}

\appendix

\section{Appendix A: Many-body Wavefunctions}

The ground state wavefunction is $ \Psi_{GS} = \ket{0} $, and the excited states with single excitation and double excitations are 
\begin{align}
\Psi_1^n(N_{site}) &= \sum_{i, j [i, j \in \binom{N_{site},\sigma = \uparrow, \downarrow}{1}] }  a_{i, j, n}  d^{v\dagger}_{i} c_{j}^{c\dagger} \ket{0} , \\ 
\Psi_2^n(N_{site}) &= \sum_{\substack{(i, j), (k, l) \\ [(i, j), (k, l) \in \binom{N_{site},\sigma = \uparrow, \downarrow}{2}]} }  a_{(i, j), (k, l), n}  d^{v\dagger}_{i} d^{v\dagger}_{j} c_{k}^{c\dagger} c_{l}^{c\dagger} \ket{0} ,
\end{align}

respectively, where $N_{site}$ is number of sites in our model, $\binom{N_{site},\sigma = \uparrow, \downarrow}{p}$ denote all possibilities of choosing $p$ element(s) from $2N_{site}$ orbitals for both spin up and down electrons, $i, j, k, l$ are indices for the site numbers, and $n$ is the state number.
The mixing from Slater determinants with other excitation numbers in our double excitation states is less than $10^{-5}$, so we neglect their contribution in our analysis.  
Since the exciton-exciton interaction involves all possible interactions and correlations between two 
e-h pairs and is a four-particle property, we need excitation states information up to double excitation, i.e., excitation with more than 2 electrons to the unoccupied manifold, leaving 2 holes in the occupied manifold, to fully describe the exciton-exciton interactions.

\section{Appendix B: Green's function descriptions}
\vspace{-0.6\baselineskip}

A conceptual link between simulations and measured spectra is provided by Green's function. The exciton-exciton interaction is fully captured by the four-point Green's function $G^{(4)}(1, 2, 3, 4; 1', 2', 3', 4')$ 
{$ = \langle 0|T\hat\psi(1)\,\hat\psi(2)\,\hat\psi(3)\,\hat\psi(4)\,
\hat\psi^\dagger(4')\,\hat\psi^\dagger(3')\,\hat\psi^\dagger(2')\,\hat\psi^\dagger(1') |0 \rangle$ where $1\equiv (\mathbf r_1, \tau_1, \sigma_1)$, etc.  It involves 2 creation and annihilation operators for the electron excitation and 2 for hole excitations.}
A correlated e-h  propagator follows $L(1,2;1’,2’) = -i \langle T \left[ \Psi_c(1) \Psi_v^\dagger(2) \Psi_v(2’) \Psi_c^\dagger(1’) \right] \rangle$, while an independent e-h propagator $L_0(1,2;1’,2’) = G_c(1,1’) \cdot G_v(2’,2)$, can be written as the direct product. 
Similarly, a correlated exciton-exciton propagator $G^{(4)}(1, 2, 3, 4; 1', 2', 3', 4') = -i \langle T \left[ X_1(1, 2) X_2^\dagger(3, 4) X_2(4’, 3') X_1^\dagger(2', 1’) \right] \rangle $. Where $X_i$ are contracted electron-hole pair operators.
For weakly interacting exciton systems, the exciton-exciton propagator can be separated as $G^{(4)}_0(1, 2, 3, 4; 1', 2', 3', 4') =L(3,4;3’,4’) L(1,2;1’,2’) $, and when the e-h pairs are also weakly interacting, it can be direct product of independent two e-h propagators:  $G^{(4)}_{0, 0}(1, 2, 3, 4; 1', 2', 3', 4') = G_c(3,3’)  G_v(4’,4)  G_c(1,1’)  G_v(2’,2) $. 
We name these three cases of double exciton states as the SEF, WEF, and EG, respectively. 
We note that this division of three phases does not rely on energetics.

\section{Appendix C: The definition of WEF and SEF phases}
\vspace{-0.6\baselineskip} 
We define and characterize the two correlated exciton phases with the help of two auxiliary models (see SM): (1) A model with only onsite $V^0$ interactions but with long-range $V^i = 0 (i>0)$, and (2) A model with all attractive interactions $V^i = 0 $. All the repulsive interactions are included for both auxiliary models. 
Model 1 represents a system with nearly independent excitons since the absence of long-range interaction suppresses the interaction between excitons, while model 2 represents a system with nearly free electron and hole quasiparticles since the absence of attractive interaction hinders the e-h binding. We denote the many-body wavefunction solved in the original system, model 1, and model 2 by: $\Psi_{f,2}^n(t) = \Psi_2^n(V^i, U^i, t)$, $\Psi_{onsite,2}^n(t) = \Psi_2^n(V^0,V^{i>0}=0, U^i, t)$, and $\Psi_{no,2}^n(t) = \Psi_2^n(V^{i}=0, U^i, t)$, respectively.
The projection $P_{onsite}^n(t) = |\braket{\Psi_{f,2}^n(t) | \Psi_{onsite,2}^n(t)}|^2$ shows how the double exciton states in system with full interactions are similar to states in a representative almost independent exciton system, and the projection $P_{no}^n(t) = |\braket{\Psi_{f,2}^n(t) | \Psi_{no,2}^n(t)}|^2$ shows how the double exciton states in system with full interactions are similar to states in a representative almost independent quasiparticle electron and hole system. The projection $P_{onsite}^n(t)$ is generally larger than $P_{no}^n(t)$ (see SM), which shows certain degrees of correlation of e-h pairs inside an exciton. 
The projections increase monotonously as the hopping $|t|$ increases, showing that larger kinetic energy suppresses the correlations. 
We choose threshold values on the two projections to classify the phases accordingly. When $P_{onsite}^n(t) < P_{Ex-Ex}$, the system is in the SEF phase, and when $P_{onsite}^n(t) > P_{Ex-Ex}$ and $P_{no}^n(t)  < P_{e-h}$, the system is in the WEF phase, and when $P_{no}^n(t) > P_{e-h}$ the system is in the EG phase. $P_{Ex-Ex}$ and $P_{e-h}$ are both chosen to be 0.995. 
The phase boundaries based on wavefunction projections are calculated for $2 \le N_{site} \le 9$, and linearly extrapolated to the thermodynamic limit (see SM), shown on the left color bar of Fig. \ref{fig:2}(b).
We find that the extrapolated phase boundaries between SEF, WEF, and EG phases are almost in the range $\sim 0.5 < \frac{t}{V^0} < 2$ in all parameter spaces we have tested. 
These phase boundaries are often located inside the SB phase[Fig. \ref{fig:2}(c)], so that in the moderate bounded regime (outside the SB and e-h plasma phases) the exciton fluid rarely exists. 
We note that the phase boundaries are dependent on the threshold values, so the phase boundaries only show a qualitative behavior.

\bibliography{main}

\end{document}